\documentclass[fleqn,twoside]{article}
\usepackage{espcrc2}

\usepackage{graphicx}
\usepackage[figuresright]{rotating}

\newcommand{\AmS}{{\protect\the\textfont2
  A\kern-.1667em\lower.5ex\hbox{M}\kern-.125emS}}

\title{Ultra-high energy cosmic rays: a probe into new physics}

\author{P. Blasi \address[Arcetri]{INAF/Osservatorio Astrofisico di Arcetri\\
Largo E. Fermi, 5 50125 Firenze, ITALY}}
       
\begin{document}

\begin{abstract}
The most energetic particles ever detected exceed $10^{20}$ eV in energy.
Their existence represents at the same time a great challenge for particle 
physics and astrophysics, and a great promise of providing us for a probe of 
the validity of the laws of Nature in extreme conditions. We review here 
the most recent data and the future perspectives for detection of cosmic 
rays at ultra-high energies, and discuss possible ways of using these data 
to test the possibility that new Physics and/or new Astrophsyics may be
awaiting around the corner.\par\vskip .3cm

\hskip 6cm
{\footnotesize \it ``When you carry out an experiment there are two possible 
outcomes:}\par
\hskip 6cm
{\footnotesize\it either you confirm the theoretical expectation, and in this 
case you made}\par
\hskip 6cm
{\footnotesize \it a measurement, or you don't, and in this case you made a
discovery''.}\par
\hskip 14cm {\small\it E. Fermi}
\vspace{1pc}
\end{abstract}

\maketitle
\section{Introduction}
The cosmic ray spectrum has been now measured over a large range of energies,
that extends in its upper part to more than $10^{20}$ eV, the so-called
{\it ultra high energy cosmic rays} (UHECRs). The quest for the
origin of these high energy particles over the whole range is still open and
represents one of the big challenges for the future. In the highest energy
end of the cosmic ray spectrum, several issues make the challenge even harder: 
a) it is hard to envision possible acceleration sites where particles with 
energy in excess of $10^{20}$ eV may be accelerated; b) even if some class of 
sources could indeed accelerate particles to the highest energies, a 
homogeneous spatial distribution of these sources would leave an imprint in the
cosmic ray spetrum, known as Greisen-Zatsepin-Kuzmin (GZK) cutoff, due to 
photopion production on the photons of the cosmic microwave background 
\cite{greisenzk66}. 
On the basis of the first issue, we would expect an {\it end of the cosmic ray
spectrum} to occur at some {\it high} energy. On the basis of the second issue,
we would expect a strong flux suppression (not really a cutoff) at about 
$5\times 10^{19}$ eV. 

Several experiments have been operating to detect the flux of UHECRs, starting
with Volcano Ranch \cite{linsley} and continuing with Haverah Park 
\cite{watson91} and Yakutsk \cite{efimov91} to the more recent 
experiments like AGASA \cite{agasa01,tak99,tak98,ha94}, Fly's Eye 
\cite{bird93,bird94,bird95} and HiRes \cite{kieda99}.

At present there is no clear indication that the cosmic ray spectrum comes 
to an end due to either one of these two reasons. More statistics of events
is however required to achieve a solid conclusion in this respect.

The physics involved in the explanation of the origin and propagation of
UHECRs needs often to be pushed to its extremes to accomodate observations. 
This transforms a problem into a precious tool to probe a territory which
is still uncharted, a New Physics which several hints tell us should exist,
but is still hidden somewhere.
In the following we describe a few directions in which this investigation
may lead. 

In \S \ref{sec:observations} we briefly summarize the status of current 
observations; in \S \ref{sec:gzk} we restate the basic issues that are 
known as GZK problem; in \S \ref{sec:NP} we illustrate two examples of 
new physics that can be investigated with the help of UHECRs, namely 
Physics close to the grand unification scale (\S \ref{sec:td}) and
possible violations of Lorentz invariance (\S \ref{sec:li}).
In \S \ref{sec:conc} we give our conclusions.

\section{Observations}\label{sec:observations}

The cosmic ray spectrum is measured from fractions of GeV to a (current)
maximum energy of $3\times 10^{20}$ eV. The spectrum above a few GeV and up to 
$\sim 10^{15}$ eV (the knee) is measured to be a power law with slope
$\sim 2.7$, while at higher energies and up to $\sim 10^{19}$ eV 
(the ankle) the spectrum has a steeper slope, of $\sim 3.1$. At energy 
larger than $10^{19}$ eV a flattening seems to be present.

The statistics of events is changing continuously: an analysis of
the ``all experiments'' statistics was carried out in \cite{uchihori00}
and found 92 events above $4\times 10^{19}$ eV. 47 events were 
detected by the AGASA experiment. A more recent analysis \cite{agasa01} 
of the AGASA data, carried out expanding the acceptance angle to $\sim 60^o$, 
has increased the number of events in this energy region to 59.

In \cite{tak99} the directions of arrival of the AGASA events 
(with zenith angle smaller than $45^o$) above 
$4\times 10^{19}$ eV were studied in detail: no appreciable departure 
from isotropy was found, with the exception of a few small scale 
anisotropies in the form of doublets and triplets of events within 
an angular scale comparable with the angular resolution of the experiments 
($\sim 2.5^o$ for AGASA). This analysis was repeated in \cite{uchihori00}
for the whole sample of events above $4\times 10^{19}$ eV, and a total of 12 
doublets and 3 triplets were found within $\sim 3^o$ angular scales. 
The attempt to associate these multiplets with different types of local 
astrophysical sources possibly clustered in the local supercluster did 
not provide evidence in that direction \cite{stanev}. 

Recently, the AGASA collaboration reported on the study of the small scale 
anisotropies in the extended sample of events with zenith angle $<60^o$: 
5 doublets (chance probability $\sim 0.1\%$) and 1 triplet (chance
probability $\sim 1\%$) were found. 

The information available on the composition of cosmic rays at the highest 
energies is quite poor. A study of the shower development was possible only 
for the Fly's Eye event \cite{bird95} and disfavors a photon primary 
\cite{halzen}. A reliable analysis of the composition is however 
possible only on statistical basis, because of the large fluctuations 
in the shower development at fixed type of primary particle.
The Fly's Eye collaboration reports evidence for a predominantly heavy 
composition at $3\times 10^{17}$ eV, with a smooth transition to light 
composition at $\sim 10^{19}$ eV. This trend was later not confirmed by AGASA 
\cite{ha94,yoda98}.
Recently in ref. \cite{zas2000} the data of the Haverah Park experiment 
on highly inclined events were re-analyzed: this new analysis results in
no more than $30\%$ of the events with energy above $10^{19}$ eV being 
consistent with photons or iron (at $95\%$ confidence level) and no
more than $55\%$ of events being photons above $4\times 10^{19}$ eV.

Recently new data have been presented by the HiRes experiment
\cite{hiresICRC}. These new results seem to be in agreement with the
presence of a GZK feature in the cosmic ray spectrum, and are therefore
in disagreement with the AGASA data. In order to quantify the discrepancy 
we ran some simulations \cite{dbo} of the statistics of events expected 
with the AGASA and HiRes exposures. In table I we report our results: the
first column gives the energy threshold used to measure the number of events
quoted in the other columns. The injection spectrum is taken as $E^{-2.4}$,
because it provides the best fit to the data points of AGASA at lower 
energies. The evolution in the luminosity of the sources is taken as $L\propto 
(1+z)^{m+3}$, where the 3 accounts for the redshift ($m=0$ 
corresponds to no intrinsic evolution). For AGASA we also considered the 
possibility that there is a systematic error of $10\%$ in the energy 
determination (column indicated as $m=0 + 10\%$ sys.). From the numbers
in the table we deduce that the AGASA data do not agree with the GZK 
prediction at the level of $2.4 \sigma$ ($3.6\sigma$) for $m=0$ ($m=4$).
On the other hand HiRes data seem to agree with the predicted flux. 

Clearly we need more data in order to finally settle the issue. 

\begin{table*}[htb]
\begin{center}
\scriptsize
\begin{tabular}{c|cc|cc|cc|cc|cc}
$E_{\rm thresh.}$  &  \multicolumn{4}{|c|}{HiRes}  & \multicolumn{6}{|c}{AGASA}\\
 &  \multicolumn{2}{c}{$m=0$}  &  \multicolumn{2}{|c|}{$m=4$}  &  \multicolumn{2}{|c|}{$m=0$}  &  \multicolumn{2}{|c|}{$m=0$ + 10\% sys.}  &  \multicolumn{2}{|c}{$m=4$}\\
\hline
$10^{19}$   & 346          &            & 346          &            & 811       &            & 811       &            & 811       &            \\
$10^{19.5}$ & $52\pm4$     &$0.7\sigma$ & $46\pm4$     &$0.7\sigma$ & $117\pm5$ &            & $122\pm5$ &            & $108\pm5$ &            \\
$10^{19.6}$ & $32\pm4$     &            & $27\pm4$     &            & $72\pm5$  &$1.7\sigma$ & $75\pm5$  &$2.4\sigma$ & $63\pm4$  &$0.3\sigma$ \\
$10^{20}$   & $1.7\pm1.7$  &$0.2\sigma$ & $1.2\pm1.6$  &$0.5\sigma$ &$3.9\pm2.2$&$2.8\sigma$ &$4.4\pm2.2$&$2.6\sigma$ &$2.9\pm2.0$&$3.6\sigma$ \\
\hline
\end{tabular}
\caption{\it Simulated events for the AGASA and HiRes exposures. The injection 
spectrum is taken as $E^{-2.4}$. The source evolution is taken proportional 
to $(1+z)^{3+m}$.}
\end{center}
\end{table*}

\section{The physical meaning of the GZK feature}\label{sec:gzk}

The puzzle of UHECRs can be summarized in the following points:

\begin{itemize}
\item {\it \underline{The production problem}:}
Acceleration mechanisms have to be pushed to their extremes in order to allow
the production of particles with energy in excess of $10^{20}$ eV.

\item {\it \underline{The large scale isotropy}:}
observations show a remarkable large scale isotropy of the arrival 
directions of UHECRs, with no correlation with local structures
(e.g. galactic disk, local supercluster, local group).

\item {\it \underline{The small scale anisotropy}:}
the small (degree) scale anisotropies, if confirmed by further upcoming 
experiments, would represent an extremely strong constraint on the type of
sources of UHECRs and on magnetic fields in the propagation volume.
At present no association has been found between the arrival directions of
UHECRs and nearby bright sources.

\item {\it \underline{The GZK feature}:}
the GZK cutoff is mainly a geometrical effect: the number of sources
within a distance that equals the pathlength for photopion production is 
far less than the sources that contribute lower energy particles, having 
much larger pathlength (comparable with the size of the universe).
The crucial point is that the cutoff is present even if plausible nearby UHECR
engines are identified.

\end{itemize}

The solution of the problem of UHECRs in its several aspects will come when
many pieces will fall into place. A careful measurement of the spectrum of 
UHECRs and of their chemical composition will play a crucial role for the
identification of the sources. In particular, the composition may be a smoking
gun in favor or against whole classes of models.

In this section we consider in some more detail the issue of the GZK
feature and the implications of its possible absence, suggested by the
AGASA data.

It is often believed that the identification of one or a class of nearby
UHECR sources would explain the observations and in particular the absence 
of the GZK feature.
This is not necessarily true. The (inverse of the) lifetime of 
a proton with energy $E$ is plotted in fig. 1 (left panel) together with
the derivative with respect to energy of the rate of energy losses $b(E)$
(right panel) [the figure has been taken from ref. \cite{bereICRC}]. 
The flux per unit solid angle at energy $E$ in some direction is proportional 
to $n_0 \lambda(E) \Phi(E)$, where $n_0$ is the density of sources (assumed 
constant), $\lambda(E)=c/((1/E)dE/dt)$ and $\Phi(E)$ is the source spectrum.
This rough estimate suggests that the ratio of detected fluxes (multiplied
as usual by $E^3$), at energies $E_1$ and $E_2$ is 
\begin{equation}
{\cal R}=
\frac{E_1^3 F(E_1)}{E_2^3 F(E_2)}\sim \frac{\lambda(E_1) \Phi(E_1) E_1^3}
{\lambda(E_2)\Phi(E_2) E_2^3} =
\frac{\lambda(E_1)}
{\lambda(E_2)} \left(\frac{E_1}{E_2}\right)^{3-\gamma},
\label{eq:ratio}
\end{equation}
where in the last term we assumed that the source spectrum is a power law
$\Phi(E)\sim E^{-\gamma}$.
If for instance one takes $E_1=10^{19}$ eV (below $E_{GZK}$) and 
$E_2=3\times 10^{20}$ eV (above $E_{GZK}$), 
from fig. 1 one obtains that ${\cal R}\sim 80$ for $\gamma=3$ and 
${\cal R}\sim 10$ for $\gamma=2.4$. The ratio ${\cal R}$ gives a rough
estimate of the suppression factor at the GZK cutoff and its dependence
on the spectrum of the source. For flat spectra ($\gamma\leq 2$) the 
cutoff is less significant, but it is 
more difficult to fit the low energy data \cite{blanton} 
(at $E\sim 10^{19}$ eV). Steeper
spectra make the GZK cutoff more evident, although they allow an easier 
fit of the low energy data.
The simple argument illustrated above can also be interpreted in an alternative
way: if there is a local overdensity of sources by a factor $\sim {\cal R}$,
the GZK cutoff is attenuated with respect to the case of homogeneous 
distribution of the sources. 
\begin{figure}[t]
  \vspace{6.0cm}
  \includegraphics{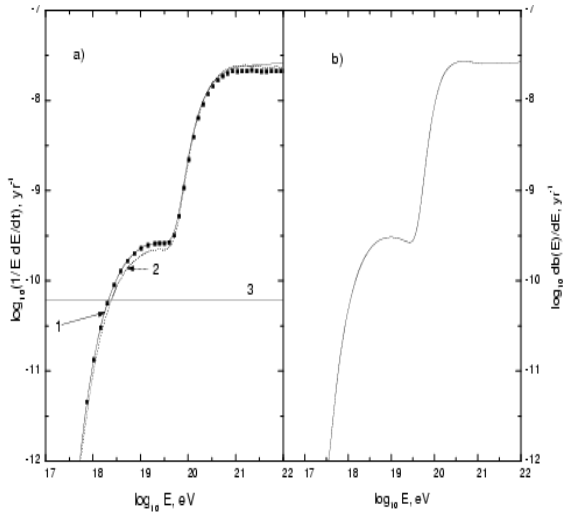}
  \caption{\it From \cite{bereICRC}. Left panel) $(1/E)dE/dt$ for a proton in 
\cite{bereICRC} (curve 1), in \cite{grigo88} (curve 2) and in \cite{stanev00}
(black squares). The curve 3 is the contribution of the red shift. Right panel)
The derivative $db(E)/dE$, with $b(E)=dE/dt$ at $z=0$.
    \label{fig1} }
\end{figure}
The question of whether we are located in such a large overdensity of 
sources was addressed, together with the propagation of UHECRs,
in \cite{blanton}. Assuming that the density of the (unknown) sources
follows the density of galaxies in large scale structure 
surveys like PSCz \cite{pscz} and Cfa2 \cite{cfa2}, the authors estimate
the local overdensity on scales of several Mpc to be of order $\sim 2$,
too small to compensate for the energy losses of particles with energy
above the threshold for photopion production. 

As discussed in \S \ref{sec:observations}, the current observational 
situation is not crystal clear in pointing toward the absence or the
presence of the GZK feature in the data. Recalling the sentence of Enrico
Fermi, quoted in the beginning of this paper, we are in a ironic situation
in which we have a $2.5 \sigma$ discovery from AGASA, and an equally 
significant measurement from the HiRes collaboration. What is absolutely clear
is the need for higher statistics experiments, such as Auger and
EUSO, that fortunately are on the way.

The bottom line of this section can be summarized in the following few points:

1) the GZK cutoff is not avoided by finding sources of UHECRs that lie 
within the pathlength of photopion production, unless these sources are
located only or predominantly nearby and are less abundant at large 
distances. 

2) To establish without any doubt the statistical significance of the
GZK feature in the cosmic ray data we need a statistics of events with energy 
$\geq 10^{20}$ eV larger than the present one. The enhanced statistics will
also make easier to study the chemical composition and to identify the sources 
of UHECRs.

\section{Probing new physics \label{sec:NP}}

UHECRs provide a laboratory where we may investigate the extremes of 
the Physics we know and possibly check whether there is something more
than that. In this paper, we discuss only two examples, one related to
the production of UHECRs (Top-down models) and the other, possible 
minuscule violations of Lorentz invariance, that might affect the 
propagation of UHECRs on cosmological distances. Many more examples may
be found but will not beb discussed here. A come comprehensive review can 
be found in \cite{sigl01}.

\subsection{Top-Down (TD) models \label{sec:td}} 

The old question of how the UHECRs are produced is as fashionable now
as it was 30 years ago. The severe requirements needed for ordinary 
acceleration processes to achieve ultra-high energies have fueled much
interest in production mechanisms that work from the top, meaning that 
particles are not accelerated but rather injected as a result of the 
decay of very massive unstable relics of the big bang. This 
may occur either as a result of decay processes of topological defects 
or by decays/annihilations of supermassive relic particles. 

Topological defects are naturally formed at phase transitions and
their existence has been proven by direct observations in several 
experiments on liquid crystals and ferromagnetic materials. 
Similar symmetry breakings at particle physics level are responsible for the
formation of cosmic topological defects [for a review see \cite{vilshe94}]. 

The fact that topological defects can generate UHECRs was first proposed in
the pioneering work in Ref. \cite{hsw87}. The general idea 
is that the stability of the defect can be locally broken by different types of
processes and result in the false vacuum, trapped within the 
defect, to fall into the real vacuum (outside universe), so that the gauge
bosons of the field trapped in the defect acquire a mass $m_X$. 
At this point, the very massive and unstable particles rapidly decay 
producing high energy particles. 

Several topological defects have been studied in the literature: ordinary
strings \cite{rana90}, superconducting strings \cite{hsw87}, 
bound states of magnetic monopoles \cite{hill83,bs95}, networks 
of monopoles and strings \cite{martin}, necklaces \cite{berevile97} and 
vortons \cite{masperi}.

The basic idea, common to all TD models, is that
the decay of a supermassive particle results in the production of a 
quark-antiquark pair that hadronizes into mesons and protons. At the source,
the composition of the produced particles is dominated by gamma rays and
neutrinos, while only about $5\%$ of the energy goes into protons. After 
propagation over cosmological distances, the relative abundance of gamma
rays and protons changes, as illustrated in fig. 2. Gamma rays become then
dominant only at extremely high energies, although an appreciable fraction 
of the composition is still made of gamma rays.
\begin{figure}[t]
  \vspace{5.0cm}
  \includegraphics{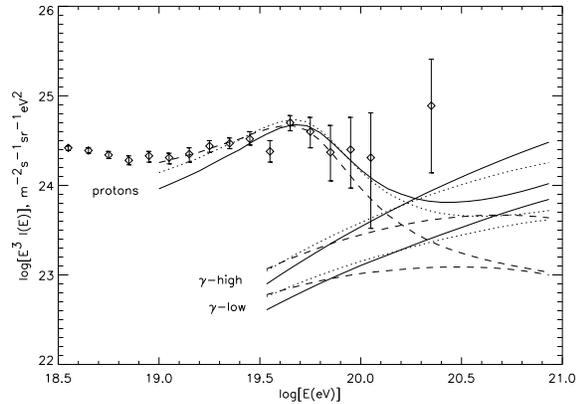}
  \caption{\it Spectrum of protons and gamma rays from topological defects 
\cite{bbv}. The two sets of curves for gamma rays are obtained for two 
different estimates of the radio background. Different curves refer to 
different values of the mass of X-particles.
    \label{fig2} }
\end{figure}
Much discussion exists on which topological defects may generate the observed
fluxes, as summarized in \cite{bbv,bhattasigl}. 

An interesting alternative to topological defects is represented by 
super-heavy (SH) relics of the big bang.
SH particles with very long lifetime can be produced in the early 
universe and generate UHECRs at present 
\cite{berekacvil97,kr98,ckr98,kt98,ktrev}. 
In the following, we will call these particles {\it X-particles}.

X-particles can be produced in the early universe through different
mechanisms. The simplest of them is the {\it gravitational production}:
particles are produced naturally in a time variable gravitational field
or indeed in a generic time variable classical field. In the gravitational
case no additional coupling is required (all particles interact 
gravitationally). If the time variable field is the inflaton field $\phi$, 
a direct coupling of the X-particles to $\phi$ is needed. 

The gravitational production of particles was first proposed in 
\cite{zelsta72}. It does not require any additional assumption
neither on the X-particles nor on cosmology. In particular inflation is 
not required a priori, and indeed it reduces the effect. It can be shown 
that at time t, gravitational production can only generate X-particles 
with mass $m_X\leq H(t)\leq  m_\phi$, where $H(t)$ is the Hubble constant
and $m_\phi$ is the inflaton mass. It was demonstrated in \cite{ckr98,kt98} 
that the fraction of the critical mass contributed by X-particles with 
$m_X\sim 10^{13}$ GeV produced gravitationally is $\Omega_X\sim 1$, with no 
additional assumption! 
In other words, cold dark matter can naturally be explained in terms of 
X-particles in this range of masses.

As mentioned in the beginning of this section, in order for X-particles 
to be useful dark matter candidates and generate UHECRs they need to be long 
lived. The gravitational coupling by itself induces a lifetime much shorter 
than the age of the universe for the range of masses which we are interested 
in. Therefore, in order to have long lifetimes, additional symmetries must be 
postulated: for instance discrete gauge symmetries can protect X-particles 
from decay, while being very weakly broken, perhaps by instanton effects 
\cite{kr98}. These effects can allow decay times larger than 
the age of the universe, as shown in \cite{hama98}.

The slow decay of X-particles produces UHECRs. The interesting feature of 
this model is that X-particles cluster in the galactic halo, as cold dark 
matter \cite{bbv}. 
Hence UHECRs are expected to be produced locally, with no absorption. As 
a consequence, the observed spectra are nearly identical to the emission 
spectra, and therefore gamma rays dominate. The very flat 
spectra and the gamma ray composition are two of the signatures.
The calculations of the expected fluxes have been performed in 
\cite{bbv,bsarkar98,blasi99}. In figure 3 we report a typical prediction 
of the spectra for this model, as derived in \cite{dick}.
\begin{figure}[t]
  \vspace{5.0cm}
  \includegraphics{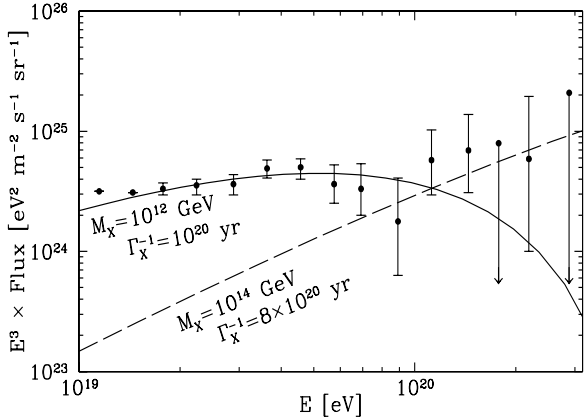}
  \caption{\it Spectra of gamma rays from the decay of SH relics. The 
parameters are chosen as indicated \cite{dick}.   
    \label{fig3} }
\end{figure}
The strongest signature of the model is the anisotropy due to the 
asymmetric position of the sun in the Galaxy 
\cite{dubo98,bbv,beremika99,medwat99}. These papers agree that the present
data are consistent with the anisotropy expected in the model of 
X-particles in the halo, due to the large uncertainties in the measurement
of anisotropy. In fact, as discussed in \S \ref{sec:observations}, 
observations at present do not show any appreciable deviation from 
isotropy, with the exception of a few degree scale anisotropies showing 
up in the form of doublets and triplets of events within an angular scale 
comparable to the resolution of the experiments. 

In TD models the presence of these multiplets of events is not well 
accomodated because of the homogeneous distribution of the topological
defects or of the SH particles. However, it was shown in \cite{blsh00}
that in the latter the presence of the multiplets may actually be accomodated.

As an alternative to the long lifetimes required by the decay of SH relics, 
a model based on their annihilation has been proposed \cite{dick}, where
however large annihilation cross-sections are required.

\subsection{Lorentz invariance: is it really a good symmetry? \label{sec:li}}

UHECRs are the particles with the largest Lorentz factor that we have been
able to measure. They are then the most powerful probe to test the validity
of Lorentz invariance (LI), one of the fundamental ideas of known Physics.
LI violations affect the propagation of UHECRs mainly because they may change 
the kinematic thresholds for photopion production, although other effects 
have also been investigated. The pioneering work in this direction was 
presented in \cite{livio}. A very general parametrization of the violation 
of LI was discussed in \cite{grillo1}, where it was shown that the threshold 
may even move to infinite energy of the ultra high energy proton, meaning that
the reaction becomes kinematically forbidden. The possible violation of Lorentz
invariance at the Planck scale has been proposed by several authors as a 
possible explanation of the detection of events with energy above $10^{20}$ 
eV. However, it is clear from the discussion in \S \ref{sec:observations} that
the current number of such events is not sufficient to justify such a 
proposal. 
Nevertheless, the shape of the spectrum of UHECRs can provide interesting 
constraints on the strength of the LI violation, if the GZK feature is 
detected by upcoming experiments. 

Recently, an interesting extension of the previous ideas on violation of LI
was discussed in \cite{ng1}. In the new approach, the violation is 
induced by the quantum fluctuations of the space-time foam: if the Planck
length is the smallest length which can be measured, then this scale also 
represents a lower limit on the error in a length measurement, and in 
particular of the De Broglie length of a particle. The induced uncertainty 
in the momentum of the particle is $\delta p \sim p^2/M_P$, so that energy
and momentum can be written as \cite{grillo2}
\begin{equation}
E={\tilde E} + \alpha \frac{{\tilde E}^2}{M_P} ~~~
p={\tilde p} + \beta \frac{{\tilde p}^2}{M_P}, 
\end{equation}
where $\alpha$ and $\beta$ are randomly distributed numbers with unit 
variance, and $\tilde E$ and $\tilde p$ are the average energy and momentum. 
Moreover the fluctuations in the metric can be written in the following form:
\begin{equation}
p_\mu g^{\mu\nu} p_\nu = E^2 - p^2 + \gamma  \frac{p^3}{M_P} = m^2,
\end{equation}
where $\gamma$ is also a random number with normal distribution and unit
variance. When applied to the process of photopion production, the effect
of the metric fluctuations is to move the threshold for photopion production
around the {\it classical} value. However, as demonstrated in \cite{grillo2},
when the reaction remains kinematically allowed, the threshold is moved 
to $\sim 10^{15}$ eV, so that on average the effect of violation of LI due
to a fluctuating metric is mostly of making low energy particles vulnerable
to photopion production \cite{grillo2} 

\section{Conclusions \label{sec:conc}}

The search for the end of the cosmic ray spectrum, started a few decades
back in time, is still ongoing and still not successful. This challenge
led us to the detection of particles with energies in excess of $10^{20}$
eV. Acceleration processes are strongly limited by energy losses and
finite size of the known acceleration regions and only some types of sources
are barely able to energize protons up to the observed energies \cite{olinto}.
Composition and anisotropy studies will be the keys to solve the mystery,
but at the cost of increasing the statistics by at least a factor of 10
compared with current experiments. Two experiments are being planned for 
the next decade or so, and will provide the characteristics necessary to 
do cosmic ray astronomy: the Auger project \cite{cronin} is currently in 
the construction stage in Argentina, while the EUSO project \cite{scarsi}
is scheduled for operation starting in 2008. 
Each one of these enterprises implies an improvement by a factor
$\sim 10$ compared with the previous one, which means a predicted 500 events
per year above $10^{20}$ eV for EUSO, if the AGASA spectrum is taken as a
template. 

Besides being the tools for ultra high energy cosmic ray astronomy, these
experiments represent a unique tool to study possible New Physics at extremely
high energies. The case of neutrino oscillations provides an example of 
the first hint of the existence of Physics beyond the Standard Model of 
Particle interactions, derived in an Astrophysics context.
It is foreseeable that the ball of particle physics, after a few decades,
could go back to the field of cosmic rays, where the first steps in that
direction were moved in the '30s.


\begin{thebibliography}{9}

\bibitem{greisenzk66} K. Greisen, Phys. Rev. Lett. {\bf 16}, 748 (1966);
G.T. Zatsepin and V.A. Kuzmin, Sov. Phys. JETP Lett. {\bf 4}, 78 (1966).

\bibitem{linsley} J. Linsley, Phys. Rev. Lett. {\bf 10}, 146 (1963).

\bibitem{watson91} 
M.A. Lawrence, R.J.O. Reid and A.A. Watson, J. Phys. G. Nucl.
Part. Phys. {\bf 17}, 773 (1991).

\bibitem{efimov91} 
N.N. Efimov {\it et al}: Ref. Proc. International Symposium on 
{\it Astrophysical Aspects of the most energetic cosmic rays}, eds M. Nagano
and F. Takahara (World Scientific, Singapore), p. 20 (1991).

\bibitem{agasa01}
M. Teshima, proceedings of TAUP 2001, Laboratori Nazionali del
Gran Sasso, L'Aquila, Sept. 8-12, 2001.

\bibitem{tak99} M. Takeda {\it et al} Astrophys. J. {\bf 522}, 225 (1999).

\bibitem{tak98} M. Takeda {\it et al}, Phys. Rev. Lett. {\bf 81}, 1163 (1998).

\bibitem{ha94} N. Hayashida {\it et al}, Phys. Rev. Lett. {\bf 73}, 3491 
(1994).

\bibitem{bird93} D.J. Bird {\it et al}, Phys. Rev. Lett. {\bf 71}, 3401 
(1993). 

\bibitem{bird94}  D.J. Bird {\it et al},
Astrophys. J. {\bf 424}, 491 (1994).

\bibitem{bird95} 
D.J. Bird {\it et al}, Astrophys. J. {\bf 441}, 144 (1995). 

\bibitem{kieda99} 
D. Kieda {\it et al}, {\it HiRes Collaboration} 1999 Proc. of 26th ICRC,
Salt Lake City, Utah.

\bibitem{uchihori00} 
Y. Uchihori, M. Nagano, M. Takeda, M. Teshima, J. Lloyd-Evans and
A.A. Watson,  Astropart. Phys. {\bf 13}, 151 (2000).

\bibitem{stanev}
T. Stanev, proceedings of the Vulcano Workshop ``Frontier
Objects in Astrophysics and Particle Physics'', Vulcano, May 21-27, 2000.

\bibitem{halzen} 
F. Halzen, R. Vazques, T. Stanev and H.S. Vankov, Astropart. 
Phys. {\bf 3}, 151 (1995).

\bibitem{yoda98} S. Yoshida and H. Dai, J. Phys. G {\bf 24}, 905 (1998).

\bibitem{zas2000}
M. Ave, J.A. Hinton, R.A. Vazquez, A.A. Watson, E. Zas 
Phys. Rev. Lett. {\bf 85}, 2244 (2000).

\bibitem{hiresICRC}
Numerous contributions have been presented by the HiRes Collaboration at the
ICRC2001.

\bibitem{dbo}
D. De Marco, P. Blasi and A.V. Olinto, in preparation.

\bibitem{bereICRC}
V. Berezinsky, A.Z. Gazizov, S.I. Grigorieva, preprint hep-ph/0107306.

\bibitem{grigo88}
V.S. Berezinsky and S.I. Grigorieva, A\&A, {\bf 199}, 1 (1988).

\bibitem{stanev00}
T. Stanev {\it et al}, Phys. Rev. {\bf D62}, 093005 (2000).

\bibitem{blanton}
M. Blanton, P. Blasi and A.V. Olinto, Astrop. Phys. {\bf 15}, 275 (2001).

\bibitem{pscz}
W. Saunders {\it et al}, preprint astro-ph/0001117.

\bibitem{cfa2}
J.P. Huchra, M.J. Geller and H.J. Corwin Jr., Astroph. J. {\bf 70} 687 (1995).

\bibitem{olinto} 
A.V. Olinto, Phys. Rep. {\bf 333}, 329 (2000).

\bibitem{bhattasigl} 
P. Bhattacharjee and G. Sigl, Phys. Rep. {\bf 327}, 109 (2000).

\bibitem{vilshe94} 
A. Vilenkin and E.P.S. Shellard, {\it Cosmic Strings and Other
Topological Defects}, Cambridge University Press, Cambridge (1994).

\bibitem{hsw87} 
C.T. Hill, D.N. Schramm and T.P. Walker, Phys. Rev. {\bf D36}, 1007 (1987).

\bibitem{rana90} 
P. Bhattacharjee and N.C. Rana, Phys. Lett. {\bf B246}, 365 (1990).

\bibitem{hill83} C.T. Hill, Nucl. Phys. {\bf B224}, 469 (1983).

\bibitem{bs95} P. Bhattacharjee and G. Sigl, Phys. Rev. {\bf D51}, 4079 (1995).

\bibitem{martin} 
V.S. Berezinsky, X. Martin and A. Vilenkin, Phys. Rev. {\bf D56}, 2024 (1997).

\bibitem{berevile97} 
V.S. Berezinsky and A. Vilenkin,  Phys. Rev. Lett. {\bf 79}, 5202 (1997).

\bibitem{masperi} 
L. Masperi and B. Silva, Astropart. Phys. {\bf 8}, 173 (1998).

\bibitem{bbv} V.S. Berezinsky, P. Blasi and A. Vilenkin, Phys. Rev. 
{\bf D58}, 103515 (1998).

\bibitem{berekacvil97} 
V.S. Berezinsky, M. Kachelriess and A. Vilenkin, Phys. Rev. Lett. 
{\bf 79}, 4302 (1997).

\bibitem{dick}
P. Blasi, R. Dick and E.W. Kolb, Astropart. Phys., in press.

\bibitem{kr98} V.A. Kuzmin and V.A. Rubakov, Yadern. Fiz. {\bf 61},  
1122 (1998).

\bibitem{ckr98} 
D.J.H. Chung, E.W. Kolb and A. Riotto, Phys. Rev. {\bf D59}, 023501 (1998).

\bibitem{kt98}  V.A. Kuzmin and I.I.Tkachev, JETP Lett. {\bf 69}, 271 (1998).

\bibitem{ktrev}  V.A. Kuzmin and I.I.Tkachev, Phys. Rep. {\bf 320}, 199 (1999).

\bibitem{zelsta72} 
Ya.B. Zeldovich and A.A Starobinsky, Soviet Phys. JETP {\bf 34}, 1159 (1972). 

\bibitem{hama98} 
K. Hamaguchi, Y. Nomura and T. Yanagida, Phys. Rev. {\bf D58}, 103503 (1998). 

\bibitem{bsarkar98} 
M. Birkel and S. Sarkar, Astropart. Phys. {\bf 9}, 297 (1998). 

\bibitem{blasi99} P. Blasi, Phys. Rev. {\bf D60}, 023514 (1999).

\bibitem{sarkar}
S. Sarkar, `COSMO-99, Third Intern. Workshop on Particle Physics and 
the Early Universe' pages 77-91, Trieste, 27 Sep-3 Oct 1999.

\bibitem{dubo98} 
S.L. Dubovsky and P.G. Tynyakov, Pis'ma Zh. Eksp. Teor. Fiz. 
{\bf 68}, 99 [JETP Lett. {\bf 68}, 107] (1998).

\bibitem{beremika99} 
V.S. Berezinsky and A. Mikhailov, Phys. Lett. {\bf B449}, 237 (1999).

\bibitem{medwat99} 
G.A. Medina-Tanco and A.A. Watson, Astropart. Phys. {\bf 12}, 25 (1999).

\bibitem{blsh00} 
P. Blasi and R.K. Sheth, R.K., Phys. Lett. {\bf B486}, 233 (2000).

\bibitem{sigl01}
G. Sigl, preprint hep-ph/0109202.

\bibitem{livio}
D.A. Kirzhnits and V.A. Chechin, Sov. Journ. Nucl. Phys. {\bf 15}, 585 (1971);
S. Coleman and S.L. Glashow, Nucl. Phys. {\bf B574}, 130 (2000);
L. Gonzales-Mestres, Nucl. Phys. B (Proc. Suppl.) {\bf 48}, 131 (1996);
G. Amelino Camelia, J. Ellis, N.E. Navromatos and S. Sarkar, Nature {\bf 393},
763 (1998); G. Amelino Camelia, J. Ellis, N.E. Navromatos and D.V. NAnopoulos,
Int. J. Mod. Phys. {\bf A12}, 607 (1997).

\bibitem{grillo1}
R. Aloisio, P. Blasi, P. Ghia and A.F. Grillo, Phys. Rev. {\bf D62}, 053010
(2000).

\bibitem{ng1}
Y.J. Ng, D.S. Lee, M.C. Oh and H. van Dam, 
Phys. Lett. {\bf B507} 236 (2001).

\bibitem{grillo2}
R. Aloisio, P. Blasi, A. Galante, P. Ghia and A.F. Grillo, submitted to Astrop.
Phys.

\bibitem{cronin} 
J.W. Cronin, Nucl. Phys. B. (Proc. Suppl.) {\bf 28B}, 213 (1992).

\bibitem{scarsi}
See web page: http://www.ifcai.pa.cnr.it/~EUSO/home.html

\end{thebibliography}
\end{document}